\documentstyle[twoside,fleqn,espcrc2,psfig]{article}

\title{Modulated spin and charge densities in cuprate superconductors}

\author{J. M. Tranquada\address{Physics Department, Brookhaven National
Laboratory, Upton, NY 11973, USA}%
}

\begin{document}

\begin{abstract}
Neutron scattering experiments have played a crucial role in characterizing the
spin and charge correlations in copper-oxide superconductors.  While the data
are often interpreted with respect to specific theories of the cuprates, an
attempt is made here to distinguish those facts that can be extracted
empirically, and the connections that can be made with minimal assumptions.
\end{abstract}

\maketitle

\section{Introduction}

It is well known that the parent compounds of the cuprate superconductors are
antiferromagnetic insulators.  When holes are doped into the ubiquitous CuO$_2$
planes, antiferromagnetic spin correlations survive and coexist with
superconductivity.  The nature of the magnetic
correlations in the superconducting phases has been the subject of continuing
controversy.  The presence of the dynamical antiferromagnetism is believed by
many to have a connection with the unusual transport properties of the
cuprates.  

Neutron scattering experiments have played a crucial role in characterizing the
spin correlations in the cuprates.  Results are often discussed in terms of
specific theories, so that particular interpretations of data can become too
closely associated with certain theories.  This is unfortunate, as there is
quite a bit of unambiguous information that one can extract directly from the
neutron scattering results.  When combined with complementary experimental data
and rather general theoretical arguments, a striking picture of the spin and
charge correlations within the CuO$_2$ planes emerges.

My plan of attack is as follows.  The implications of the modulated charge and
spin order observed in La$_{1.6-x}$Nd$_{0.4}$Sr$_x$CuO$_4$ will be discussed
first.  Next, comparisions with La$_{2-x}$Sr$_x$CuO$_4$, and then
YBa$_2$Cu$_3$O$_{6+x}$ will be made.  A consideration of the effects of
Zn-doping follows.  I conclude with a brief apology to theorists.

\section{La$_{1.6-x}$Nd$_{0.4}$Sr$_x$CuO$_4$}

The variation of the superconducting transition temperature, $T_c$, vs.\ $x$ in
La$_{1.6-x}$Nd$_{0.4}$Sr$_x$CuO$_4$ is similar to that in
La$_{2-x}$Sr$_x$CuO$_4$, except that there is a strong depression of $T_c$ for
$x\approx\frac18$ \cite{craw91}.  Neutron diffraction measurements
\cite{tran95a,tran97a} on single crystals with
$x=0.12$ have demonstrated the existence of two kinds of incommensurate
superlattice peaks, as illustrated in Fig.~1(a).  Magnetic peaks are displaced
from the antiferromagnetic wave vector $(\frac12,\frac12,0)$ by
$(\pm\epsilon,0,0)$ and $(0,\pm\epsilon,0)$, with $\epsilon=0.12$. (Wave vectors
are specified in reciprocal lattice units, based on a real-space unit cell with
axes {\bf a} and {\bf b} parallel to the Cu-O bonds.)   Superlattice peaks split
by an amount
$2\epsilon$ about fundamental Bragg points provide evidence for charge ordering.
(The charge order is detected indirectly through nuclear displacements, but the
same is generally true in x-ray diffraction studies of charge order, as well.)
The charge-order peaks appear at a higher temperature than the magnetic peaks,
as has recently been confirmed by x-ray scattering measurements with 100~keV
x-rays \cite{vonz97}.

\begin{figure}[t]
\centerline{
\psfig{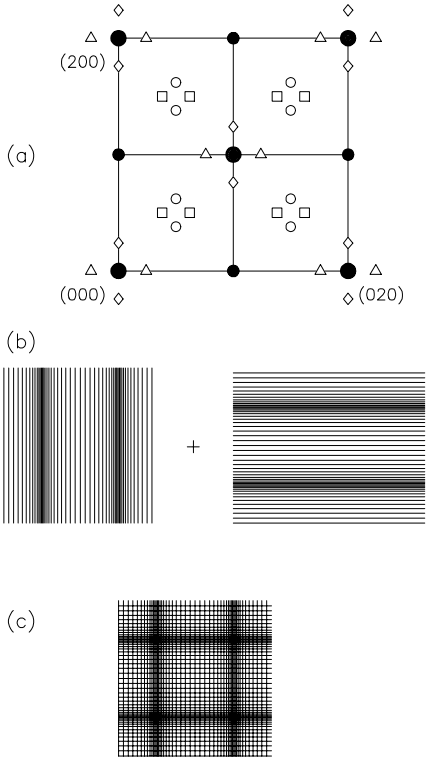}}
\caption{(a) Diagram of the $(hk0)$ zone of reciprocal space.  Filled circles:
Bragg points of the unmodulated lattice; open circles and squares: magnetic
superlattice peaks; diamonds and triangles: charge-order superlattice peaks. 
(b) and (c) illustrate real-space alternatives for the modulations: (b) twin
domains and  (c) 2{\bf Q} structure.}
\end{figure} 

Let us consider just the magnetic peaks for a moment.  The appearance of peaks
split in 2 orthogonal directions suggests 2 possibilities [see Fig.~1(b,c)]: 1)
there are 2 types of twin domains, each with a unique modulation wave vector, or
2) there is a single type of domain with a superposition of 2 orthogonal
modulations.  In the latter case, one would expect to see extra magnetic peaks
with splittings such as $\pm(\epsilon,\epsilon,0)$, whereas Shamoto {\it et al.}
\cite{sham92} demonstrated the absence of intensity at such positions.  Besides
this, there are other arguments that favor case (1).  First of all, the
individual CuO$_2$ layers in the low-temperature phase of
La$_{1.48}$Nd$_{0.4}$Sr$_{0.12}$CuO$_4$ have 2-fold, but not 4-fold, symmetry,
which would certainly be compatible with a uniaxial modulation.  The
symmetry-lowering distortion rotates by 90$^\circ$ from one layer to the next,
thus forcing equal populations of the two types of domains. Secondly, a similar
charge and spin ordering in hole-doped La$_2$NiO$_4$ has been shown to have a
single modulation wave vector
\cite{tran95b}.  Finally, singly-modulated phases are quite common in nature
\cite{seul95}, whereas doubly-modulated phases are not.

Next, there is the question of the type of spin modulation.  In general, there
could be a modulation of the spin direction, the spin amplitude, or a
combination of the two.  The existence of the charge modulation implies that
there must be a modulation of the spin amplitude.  Also, the modulation in the
case of hole-doped La$_2$NiO$_4$ appears to involve only the spin amplitude
\cite{tran95b}.  Thus, there must be a modulation of the spin amplitude, although
a component involving spin rotation is not ruled out.

Considering the spin- and charge-density modulations together, missing pieces
of information are the phases of the modulations with respect to the lattice. 
It is not clear that the absolute phases are of crucial importance.  Of
greater interest is the relative phase between the two modulations.  Given
that the parent compound is an antiferromagnet, it seems most likely that the
maxima of the charge-density modulation (corresponding to the greatest density
of dopant-induced holes) should be aligned with the nodes of the magnetic
modulation.  In fact, it is difficult to think of a scenario in which this
would not be the case.

To determine the significance of the modulations, it is necessary to evaluate
their amplitudes.  Of perhaps greatest interest is the amplitude of the charge
modulation.  One would like to know what fraction of the average hole density
is spatially modulated: all or a tiny part?  Unfortunately, neutron diffraction
cannot provide any direct information on this matter.  Instead, we must focus
on the amplitude of the spin modulation.  Here the issue is complicated by the
fact that zero-point spin fluctuations can have a significant impact on the
amplitude.  An estimate from measurements on an $x=0.12$ crystal, taking into
account the lack of long-range order, yielded an amplitude of
$0.10\pm0.03$~$\mu_B$ per Cu ion \cite{tran95a}.  Analysis of
muon-spin-relaxation ($\mu$SR) measurements on a similar crystal gives the
somewhat larger value of $\approx0.3$~$\mu_B$ \cite{nach97}.  This latter value
and the observed transition temperature ($\approx30$~K from $\mu$SR) appear to be
consistent with the empirical curve obtained when the staggered magnetization
for a series of quasi-1D and 2D antiferromagnetic insulators is plotted versus
$T_N/J$ \cite{nach97}, where $T_N$ is the N\'eel temperature and $J$ is the
superexchange energy.  Similarly, the amplitude of the spin-density modulation
in La$_2$NiO$_{4.125}$ is $>80\%$ of that found in antiferromagnetic
La$_2$NiO$_4$.  The substantial spin-density amplitudes in these systems
indicates a significant modulation of the hole density.

Any deviation from a sinusoidal modulation would be reflected in the appearance
of finite intensity at higher-harmonic superlattice positions.  For example,
one would expect magnetic scattering to appear at positions separated by
$3\epsilon$ from the antiferromagnetic point.  Such features have been measured
in La$_2$NiO$_{4.133}$ \cite{woch97}.  In that case, although the intensity
pattern of the harmonics is quantitatively consistent with a model of relatively
squared-off magnetic domains, the strongest higher harmonic (the $3^{\rm rd}$)
is only 1.5\%\ of the
$1^{\rm st}$ \cite{woch97}.  In the case of the cuprates, such a weak harmonic
would be quite difficult to detect above the background.  Furthermore, transverse
fluctuations of the spin and charge densities would likely damp out the
harmonics.

Finally, we need to consider what is driving the modulated order. 
Experimentally, it is observed that the charge-order peaks appear at a higher
temperature than the magnetic peaks \cite{tran95a}.  The significance
of this result can be evaluated in terms of a Landau model, which takes into
account only the symmetries of the system \cite{zach97}.  Comparison with the
phase diagram of the Landau model \cite{zach97} indicates that it is the
charge density that is driving the order.  Charge-driven ordering is also
observed in hole-doped La$_2$NiO$_4$ \cite{chen93,sach95,woch97}.
(If the driving force were associated with the spin density, then the charge and
spin ordering temperatures would have to be identical, as found in the case of
Cr \cite{pynn76}.)  Furthermore, the order of the transitions corresponds to
the region of the phase diagram where only an amplitude-modulated spin
density is allowed. 

To summarize, the following picture of the charge and spin modulations emerges
from an examination of the experimental data, especially when combined with
simple theoretical arguments and analogies to related compounds.   The free
energy associated with the dopant-induced hole density drives the ordering,
resulting in a substantial real-space modulation of the hole density.  Once the
hole density has ordered, the Cu spins in the hole-poor regions can order in a
manner that is locally antiferromagnetic, but which flips its phase by $\pi$
on crossing a maximum of the hole-density.

\section{La$_{2-x}$Sr$_x$CuO$_4$}

The cuprate system in which incommensurate scattering was first observed is, of
course, La$_{2-x}$Sr$_x$CuO$_4$ \cite{cheo91}.  Although in this
case the reported scattering is essentially all inelastic, for a given Sr content
$x$ the peak splitting $\epsilon$ \cite{yama97a} is essentially identical to that
found in the Nd-doped case \cite{tran97a} (see Fig.~2).  Superconductivity
coexists with the incommensurate spin correlations in both systems
\cite{tran97a,oste97}, but $T_c$ is strongly depressed when the correlations
have a static component.  Given the similar {\bf Q} dependence of the magnetic
scattering and the associated superconductivity in the two systems, it seems
clear that the spin correlations must have the same fundamental nature.  It
follows that, since the modulated antiferromagnetism in the Nd-doped system is
driven by the spatially oscillating hole density, a dynamic charge modulation is
implied in the case of La$_{2-x}$Sr$_x$CuO$_4$.

A wider range of Sr concentrations has been studied in the system without Nd,
and the variation of $\epsilon$ with $x$ shows an interesting trend.  For
$x>0.05$, $\epsilon$ initially increases linearly with $x$ before saturating
for $x>1/8$ \cite{yama97a,peti97}.  Since $\epsilon$ is proportional to
the inverse of the charge modulation period, the $\epsilon\sim x$ behavior
suggests that the amplitude of the charge modulation remains constant while the
period decreases with increasing $x$.  (A very similar trend is observed in
hole-doped La$_2$NiO$_4$
\cite{sach95}.)  This is consistent with an effective segregation of the
doped holes, such that an initially antiferromagnetic CuO$_2$ plane is broken up
into antiferromagnetic strips, with the strips becoming narrower as the hole
density increases.  (At small $x$, the pinning of holes by the randomly
distributed dopants dominates \cite{good97} and inhibits any well-defined
periodic charge modulation.  The effect of pinning by Sr ions is even greater
in La$_{2-x}$Sr$_x$NiO$_4$ \cite{sach95}.)  The saturation of $\epsilon$ for
$x>1/8$ clearly indicates that the period of the modulation remains constant in
that region, even though the charge density continues to increase.  

\begin{figure}[t]
\centerline{
\psfig{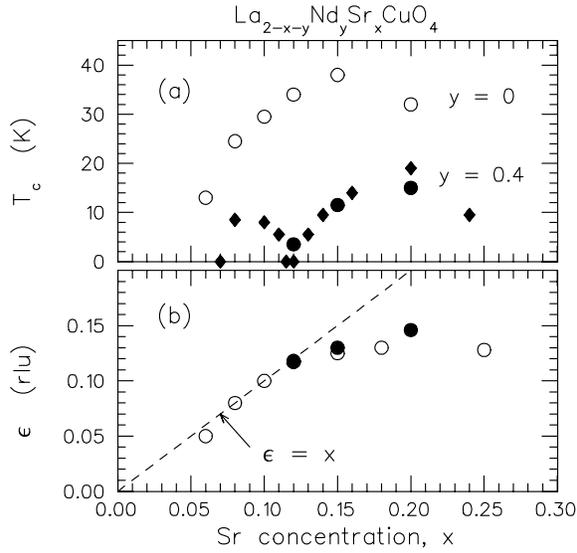}}
\caption{Results for $T_c$ and $\epsilon$ as a function of $x$ in
La$_{2-x-y}$Nd$_y$Sr$_x$CuO$_4$.  Open circles: $y=0$
\protect\cite{yama97a}; filled circles (diamonds): single-crystal
(ceramic) samples with $y=0.4$ \protect\cite{tran97a}.}
\end{figure} 

Are there big differences between samples with and without Nd doping?  Not
really.  The main effect of the Nd, which has the same valence as La, is
associated with its small ionic radius relative to Sr.  Its presence induces a
change from the usual low-temperature-orthorhombic (LTO) phase to the
low-temperature-tetragonal (LTT) phase below about 70~K \cite{craw91}, similar
to the transition first observed in La$_{1.88}$Ba$_{0.12}$CuO$_4$
\cite{axe89}.  The difference between the two structures involves a subtle
variation in the tilt pattern of CuO$_6$ octahedra.  The atomic displacement
pattern in the LTT phase is compatible with a coupling to the charge
modulation, and hence with static charge order.  In La$_{2-x}$Sr$_x$CuO$_4$,
the phonon associated with the same displacement pattern is quite soft
($\hbar\omega\approx1.5$~meV \cite{lee96}), so that it is possible that a
charge modulation, though not static, could be slowly fluctuating with the
lattice vibrations.  

The picture of antiferromagnetic strips interacting weakly across hole-rich
domain walls suggests that the spin excitations would be modified in a modest
way relative to those in the parent compound.  Indeed, inelastic neutron
scattering measurements on stripe-ordered La$_2$NiO$_{4.133}$ indicate the
low-energy spin excitations disperse away from the incommensurate peak
positions with an effective spin-wave velocity that is 60\%\ of that in undoped
La$_2$NiO$_4$ \cite{tran97c}.  Such a renormalization can be understood in
large part as occuring due to the reduction in the average number of
nearest-neighbor spins to which there is a strong coupling $J$, with no
significant change in $J$.  Inelastic measurements on La$_{2-x}$Sr$_x$CuO$_4$
\cite{cheo91,mats94,hayd96a} appear to be compatible with such a picture
\cite{tran97d}.  A reduced dispersion, limited by a finite correlation length,
is observed at low energies
\cite{cheo91,mats94}, while high-energy measurements indicate that the
maximum excitation frequency is reduced by about 20\%\ relative to
La$_2$CuO$_4$ \cite{hayd96a}.

\section{YBa$_2$Cu$_3$O$_{6+x}$}

In YBa$_2$Cu$_3$O$_{6+x}$, the antiferromagnetic spin correlations become
completely dynamic for $x\ge0.5$ \cite{kief89,bour95}.  Although most
neutron scattering studies have detected a dynamical spin susceptibility that
is peaked commensurately at the antiferromagnetic wave vector, the {\bf Q}
width of the scattering measured for $\hbar\omega<30$~meV increases roughly
linearly with $x$ \cite{bour95}, in a manner similar to the variation
of $\epsilon$ with $x$ in La$_{2-x}$Sr$_x$CuO$_4$.  The {\bf Q} dependence of
the line shape for the low-energy range and $x\sim0.5$--0.6 has been shown to
be more complex than a simple commensurately-centered gaussian
\cite{ster94}.  Recently it was pointed out that the latter results can
be modelled in a fashion similar to the incommensurate excitations in
La$_{2-x}$Sr$_x$CuO$_4$ if one allows for a shorter spin-spin correlation
length \cite{tran97d}.

In a new study of a crystal of YBa$_2$Cu$_3$O$_{6.5}$ with $T_c=52$~K, Bourges
{\it et al.} \cite{bour97} have extended inelastic scattering measurements up to
excitation energies of 120~meV.  In this good superconductor they have detected
distinct spin excitations that are very similar to the acoustic and optic
spin-wave modes observed previously in antiferromagnetic YBa$_2$Cu$_3$O$_{6.2}$
\cite{rezn96}.  Furthermore, at energies $>30$~meV the excitations show
an apparent dispersion with a velocity $\sim65$\%\ of that found in the
antiferromagnetic.  This renormalization is quite similar to that found in the
doped-nickelate case \cite{tran97c}.  The close connection between the spin
excitations in the superconducting and antiferromagnetic YBa$_2$Cu$_3$O$_{6+x}$
suggests that they are caused by similar clusters of antiferromagnetically
correlated Cu spins, and the existence of such clusters in the superconducting
sample requires a spatial segregation of the doped holes.

Dai, Mook, and Do\u gan \cite{dai97} have just reported evidence for
incommensurate magnetic fluctuations in YBa$_2$Cu$_3$O$_{6.6}$ with
$T_c=62.7$~K.  The incommensurability is observed in both energy-integrated
scans and constant-$E$ scans at $\hbar\omega=24$ and 27~meV.  It is suggested
that the modulation direction is parallel to the antiferromagnetic wave vector
(i.e., rotated by 45$^\circ$ relative to the La$_{2-x}$Sr$_x$CuO$_4$ case). 
These results appear to provide a very direct connection between the spin
correlations in superconducting YBa$_2$Cu$_3$O$_{6+x}$ and
La$_{2-x}$Sr$_x$CuO$_4$.

\section{Zn Doping}

So far, I have presented arguments that neutron scattering studies indicate an
intrinsic segregation of doped holes in the form of a periodic density
modulation within the CuO$_2$ planes.  If this interpretation is correct, then
one might expect that, besides special modifications of the crystal structure,
there would be other ways to pin the charge and spin modulations.  Indeed, light
doping with Zn appears to be another such way.

Koike {\it et al.} \cite{koik92} have shown that sustitution of 1\%\ Zn for Cu in
La$_{2-x}$Sr$_x$CuO$_4$ is sufficient to cause a drastic dip in $T_c$ as a
function of $x$, with the minimum occurring at $x=0.115$.  The resulting phase
diagram looks reminiscent of that for Nd-doped samples \cite{craw91}, except
for the lack of evidence for a change in lattice distortion.  New neutron
scattering results on a crystal of
La$_{1.86}$Sr$_{0.14}$Cu$_{0.988}$Zn$_{0.012}$O$_{4-\delta}$ with $T_c=19$~K by
Hirota {\it et al.} \cite{hiro97} show that the Zn does not modify the
incommensurability of the magnetic scattering, but does shift spectral weight
to lower energies and induces a sharp (in {\bf Q}) elastic component that grows
below $\sim30$~K.  The coexistence of superconductivity and static modulated
spin order is consistent with observations in Nd-doped samples \cite{tran97a}.

Zn-doped YBa$_2$Cu$_3$O$_{6+x}$ samples with a range of oxygen concentrations
have also been studied by neutron scattering \cite{kaku93}.  The
results are qualitatively similar to the 214 case: Zn-doping does not modify
the {\bf Q} dependence of the scattering, but it shifts spectral weight from
higher energies to the low-energy region.

It is rather interesting to note that the effect of Zn on the superconducting
cuprates is empirically quite similar to the effect of Zn-doping on spin-ladder
and spin-Peierls compounds \cite{azum97,mart97}.  Both of the latter systems,
without Zn, have energy gaps in their spin excitations.  Very small amounts of
Zn cause the coexistence of excitations within the gap and even local
antiferromagnetic order.

\section{Conclusion}

In this paper I have attempted to present a consistent interpretation of
neutron scattering measurements on a range of systems, one that I believe is
suggested directly by the data.  I hope that I have persuaded the reader that 
La$_{1.6-x}$Nd$_{0.4}$Sr$_x$CuO$_4$, rather than being a pebble in the 
collective shoe on the long march toward an understanding of the cuprates, is
actually a Rosetta stone for deciphering the experiments.  Specific microscopic
theories for the cuprates have been deliberately ignored here; hopefully, any
theorists who read this paper will feel equitably treated.  Of course, finding
the connection between the strong electronic correlations indicated by
experiments, on the one hand, and superconductivity, on the other, is entirely
the province of theorists, from whom, undoubtedly, a great deal of
``discussion'' will continue to be heard.

\bigskip

I have benefited from interactions with many experimental collaborators,
including J. D. Axe, D. J. Buttrey, G. Shirane, N. Ichikawa, S. Uchida, and P.
Wochner.  I am also grateful for frequent discussions with V. J. Emery. This work
is supported by Contract No. DE-AC02-76CH00016, Division of Materials Sciences,
U.S. Department of Energy.


\end{document}